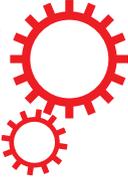

# Cooperation in the spatial prisoner's dilemma game with probabilistic abstention

Marcos Cardinot[1], Josephine Griffith[1], Colm O'Riordan[1] & Matjaž Perc [2,3]

Research has shown that the addition of abstention as an option transforms social dilemmas to rock-paper-scissor type games, where defectors dominate cooperators, cooperators dominate abstainers (loners), and abstainers (loners), in turn, dominate defectors. In this way, abstention can sustain cooperation even under adverse conditions, although defection also persists due to cyclic dominance. However, to abstain or to act as a loner has, to date, always been considered as an independent, third strategy to complement traditional cooperation and defection. Here we consider probabilistic abstention, where each player is assigned a probability to abstain in a particular instance of the game. In the two limiting cases, the studied game reverts to the prisoner's dilemma game without loners or to the optional prisoner's dilemma game. For intermediate probabilities, we have a new hybrid game, which turns out to be most favorable for the successful evolution of cooperation. We hope this novel hybrid game provides a more realistic view of the dilemma of optional/voluntary participation.

Over the last decades, the prisoner's dilemma game has been adopted in a variety of studies which seek to explore and resolve the dilemma of cooperation[1–3]. These studies include the use of the network reciprocity mechanism[4], which has gained increasing attention for its support of cooperative behaviour. In this mechanism, each agent is represented as a node in the network (graph) and is constrained to interact only with its neighbours, which are linked by edges in the network[5,6]. Research concerning network reciprocity includes the use of different topologies such as lattices[7], scale-free graphs[8–10], small-world graphs[11–13], cycle graphs[14], multilayer networks[15–17] and bipartite graphs[18,19] which have a considerable impact on the evolution of cooperation, and also favour the formation of different patterns and phenomena[20,21]. Moreover, approaches adopting coevolutionary networks, where both game strategies and the network itself are subject to evolution have also been investigated[22–31].

In essence, evolutionary game theory and its most-often used game, the prisoner's dilemma (PD) game, provides a simple and powerful framework to study the conflict between choices that are beneficial to an individual and those that are good for the whole community. The game is played by pairs of agents, who simultaneously decide to either cooperate (C) or defect (D), receiving a payoff associated with their pairwise interaction as follows: $R$ for mutual cooperation, $P$ for mutual defection, $S$ for cooperating with a defector and $T$ for successfully defecting a cooperator. The dilemma holds when $T > R > P > S$[32]. In addition to theoretical research, there is also a lot of work using experimental games. The experimental prisoner's dilemma has been used by several researchers to find mechanisms to promote cooperative behaviour, including the benefit-to-cost ratio of cooperation[33], group size[34,35], dynamic spatial structure[36,37], just to name a few examples.

Despite the overwhelming amount of scenarios that can be described as a PD game, it has been discussed that in many scenarios agents' interactions are not compulsory, and in those cases, the PD game would not be suitable. Thus, extensions of this game such as the optional prisoner's dilemma (OPD) game, also known as the prisoner's dilemma game with voluntary participation, have been explored in order to allow agents to abstain from a game interaction, that is, do not play the game and receive the so-called loner's payoff ($L$), which is the same regardless of the other agent's strategy (i.e., if either one or both agents abstain, both agents will get $L$). The dilemma is maintained when $T > R > L > P > S$[38,39]. Studies reveal that the concept of abstaining can lead to entirely different outcomes and eventually help cooperators to avoid exploitation from defectors[40–57]. Of relevance for our research is also the literature on games with an exit strategy. For example, research has been done on the dictator game with an exit strategy[58,59].

[1]Discipline of Information Technology, National University of Ireland, Galway, Ireland. [2]Faculty of Natural Sciences and Mathematics, University of Maribor, Koroška cesta 160, SI-2000, Maribor, Slovenia. [3]School of Electronic and Information Engineering, Beihang University, Beijing, 100191, P.R. China. Correspondence and requests for materials should be addressed to M.C. (email: marcos.cardinot@nuigalway.ie)

However, we believe that in many situations involving voluntary participation, such as in human interactions, the use of abstention as a pure strategy may not be ideal to capture the social dilemma. In reality, depending on the context and the type of social relationships we are modelling, abstention can also mean laziness, shyness or lack of proactivity, and all those emotions, feelings or characteristics may exist within a certain range. Thus, we propose that in a round of interactions, some agents might be interested in interacting with all of its neighbours (i.e., never abstain), while others may be willing to interact with only a few of them and abstain from interacting with others. To give another example, in the context of a poll of a number of individuals, there might be some who vote and others who do not. In the latter case, considering all the non-voters as abstainers might be too simplistic. In reality, there might be some who abstain because they do not have a view at all and those who occasionally abstain from convenience, lack of interest or because of some external event. In this way, we believe that abstention should be seen and explored as an extra attribute of each agent, and not as a pure strategy.

Given this motivation, in this paper, we introduce a prisoner's dilemma with probabilistic abstention (PDPA), which is a hybrid of two well-known games in evolutionary game theory: the PD and the OPD game (also known as the PD game with voluntary participation). As occurs in the PD game, in the hybrid game each agent can choose either to cooperate or defect. The only difference is that in the PDPA game, in addition to the game strategy, each agent is defined by a value $\alpha = [0, 1]$ to denote a probability of abstaining from any interaction.

This work aims to investigate the differences between the PDPA game and the classic PD and OPD games. A number of Monte Carlo simulations are performed to investigate the effects of $\alpha$ in the evolution of cooperation. In order to have a more complete analysis of the evolutionary dynamics, both synchronous and asynchronous updating rules[60–62] are explored.

## Results

In order to increase the understanding of the outcomes associated with the hybrid game proposed in this paper (i.e., the prisoner's dilemma game with probabilistic abstention – PDPA), in the following experiments we adopt $\varepsilon = (1-s)(1-\alpha)$ to denote the effective cooperation rate of an agent, where $s = \{0, 1\}$ and $\alpha = [0, 1]$ correspond to the agent's strategy and its probability of abstaining from a game interaction respectively. Note that here $s = 0$ means cooperator, $s = 1$ means defector, $\alpha = 0$ indicates that the agent never abstains and $\alpha = 1$ indicates that the agent always abstains. In this way, we can have two types of agents for each strategy: the pure-cooperators and the pure-defectors (i.e., the agents who always play the game, $\alpha = 0$); and the agents who sporadically play the game (i.e., the sporadic-cooperators and sporadic-defectors, $0 < \alpha < 1$). Thus, the value of $\varepsilon$ is very important to easily distinguish between a cooperator who always abstains (i.e., $\{s=0, \alpha=1\} \Rightarrow \varepsilon = 0$), from the sporadic-cooperators (i.e., $\{s=0, \alpha=(0,1)\} \Rightarrow \varepsilon > 0$), and the pure-cooperators (i.e., $\{s=0, \alpha=0\} \Rightarrow \varepsilon = 1$).

We start by comparing the outcomes of the PDPA game with those obtained for the classic prisoner's dilemma (PD) and optional prisoner's dilemma (OPD) games for both synchronous and asynchronous updating rules (Fig. 1). We test a number of randomly initialized populations of agents playing the PDPA game with three different setups:

- $\alpha = 0$ for all agents (equivalent to the PD game);
- $\alpha$ is either 0 or 1 with equal probability (equivalent to the OPD game);
- $\alpha = [0, 1]$ uniformly distributed.

For all setups, we investigate the relationship between the fraction of effective cooperation $\varepsilon$ and the probability of abstaining $\alpha$ for different values of the temptation to defect $T$ and the loner's payoff $L$.

As shown in Fig. 1, for the synchronous rule, it is possible to observe that the PDPA sustains higher levels of cooperation even for large values of the temptation to defect $T$. The difference between the outcomes of the synchronous and asynchronous versions in the classic games occur as expected: cooperation has more chance of surviving when the updating rules are synchronous, with less stochasticity and more awareness of the neighbourhood's behaviour, i.e., the agent knows who is the best player in its neighbourhood. Surprisingly, results indicate that when the PDPA is considered, this enhancement also holds for the asynchronous updating model, which is a well-known adverse scenario for both classic games[62]. In general, it is clear that irrespective of the updating rule, the PDPA game is most beneficial for the evolution of cooperation. Moreover, when comparing the OPD with the PDPA game, we see a correlation between their levels of abstention, showing that abstention may act as an important mechanism to maintain cooperation and avoid defector's dominance, which is also supported by Fig. 2, which features the time course of the fraction of $\varepsilon$ and $\alpha$ for both updating rules of agents playing the PDPA game for three values of the temptation to defect (i.e., $T = \{1.1, 1.4, 1.9\}$).

Given the nature of the classic PD and OPD games, it is known that in a well-mixed population, defection and abstention are usually the dominant strategies respectively. As discussed in previous work[22,63], this happens because cooperators need to form clusters to be able to protect themselves against exploitation from defectors, and if we consider a randomly initialized population, it takes a few steps for cooperators to cluster. Meanwhile, the defection rate increases quickly in the initial steps until the agents reach a stage where defectors have more chance of finding another defector than a cooperator. Consequently, defection starts to be a bad strategy and if abstention is an option, the agents prefer to abstain; otherwise defectors will hardly become cooperators as they do not have the incentive to change their strategies. Interestingly, as shown in Fig. 2, a similar pattern can be observed in the PDPA game, i.e., in the initial steps, the rate of pure-defectors ($\alpha = 0$) increases more quickly, causing the sporadic-cooperators ($\alpha > 0$) to have a better performance. Then, with the increase of the pure-defectors, sporadic-defectors ($\alpha > 0$) start to be a better choice. At this point, with less pure-defectors in the population, cooperators with smaller values of $\alpha$ start to perform better, producing a wave towards the decrease of $\alpha$. This simple mechanism explains the initial bell-shaped curve in the average fraction of $\alpha$ in Fig. 2.

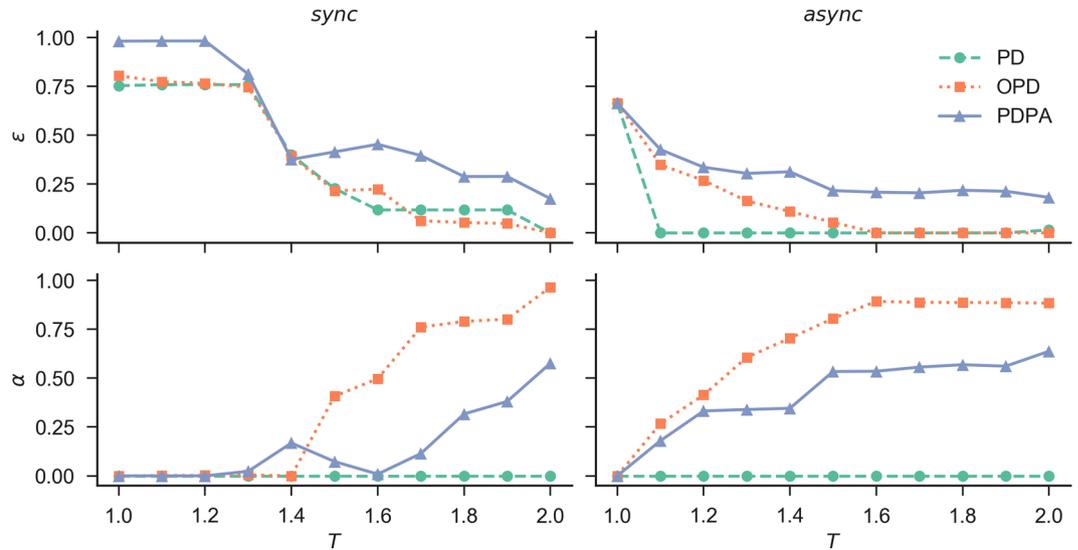

**Figure 1.** Comparing the average fractions of $\alpha$ (i.e., the probability of abstaining frequency) and $\varepsilon$ (i.e., the effective cooperation frequency — $\varepsilon = (1-s)(1-\alpha)$, where $s$ denotes the agent's strategy) for different values of the temptation to defect $T$. The results are obtained by averaging 100 independent runs at the stationary state (after $10^5$ Monte Carlo steps) of the classic prisoner's dilemma (PD), the optional prisoner's dilemma (OPD) and the hybrid of them, i.e., prisoner's dilemma with probabilistic abstention (PDPA). All games are tested in both synchronous and asynchronous updating fashions with a regular square lattice grid populated with $N = 102 \times 102$ agents, for a fixed reward for mutual cooperation $R = 1$, punishment for mutual defection $P = 0$, sucker's payoff $S = 0$ and loner's payoff $L = 0.4$.

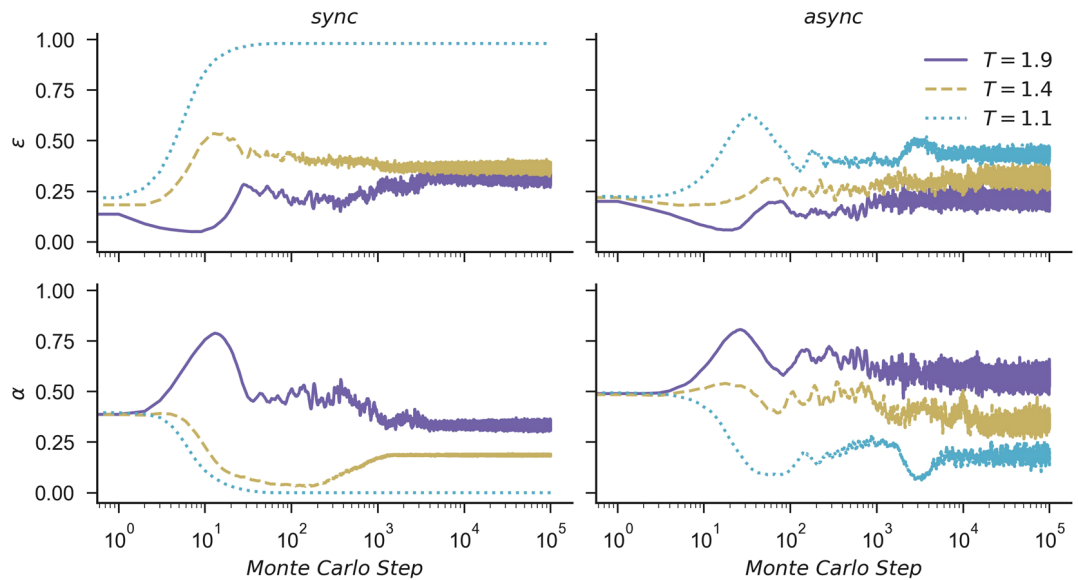

**Figure 2.** Time course of the effective cooperation $\varepsilon$ and the probability of abstaining $\alpha$ for different values of the temptation to defect $T$. All curves refer to the prisoner's dilemma with probabilistic abstention (PDPA) game for a regular square lattice grid populated with $N = 102 \times 102$ agents, for a fixed reward for mutual cooperation $R = 1$, punishment for mutual defection $P = 0$, sucker's payoff $S = 0$ and loner's payoff $L = 0.4$.

In order to further investigate the results obtained for the PDPA game, some typical distributions of the strategies, probability of abstaining $\alpha$, and the effective cooperation rate $\varepsilon$ are shown in Fig. 3. In addition to the similarities with the classic games, other interesting phenomena can be observed in the PDPA game, such as robust coexistence of cooperation and defection for different values of $T$ and $L$. Results show that agents who always refuse to interact ($\alpha = 1$) are wiped out in most scenarios when $T < 1.9$ and $L < 0.8$. That is, agents who interact at least once will usually have a better performance. Moreover, it was observed that irrespective of the high heterogeneity of values of $\alpha$ in the initialization, the population usually converges to two values of $\alpha$ for the synchronous model, and three distinct values of $\alpha$ for the asynchronous model. However, the higher heterogeneity of states in the initial steps plays a key role in increasing the performance of cooperators in the PDPA game. This happens because the intermediate values of $\alpha$ help to reduce the exposure of cooperators to the risk of being exploited by defectors too quickly.

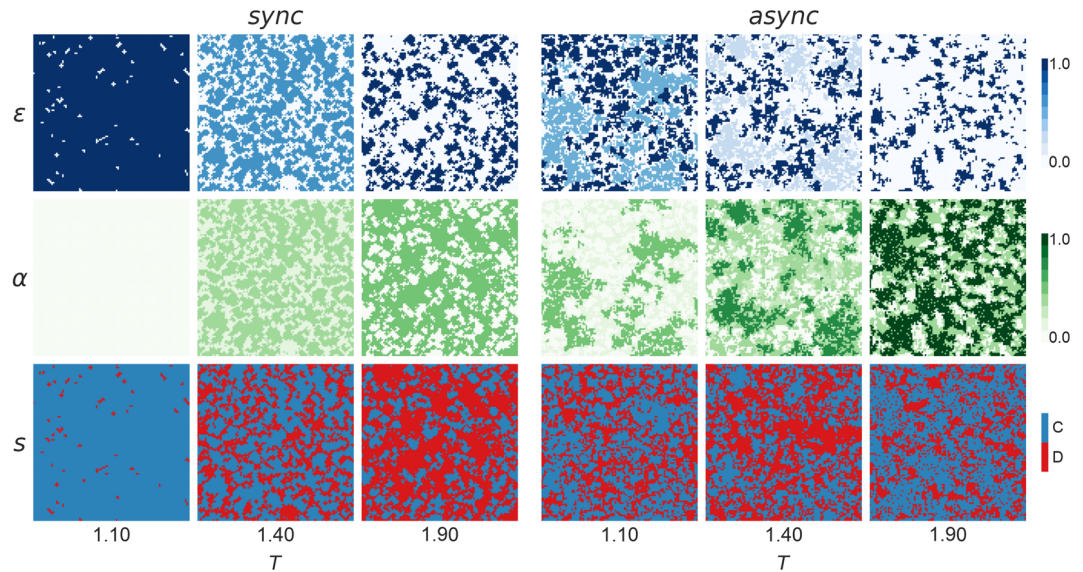

**Figure 3.** Typical distributions of the effective cooperation $\varepsilon$, probability of abstaining $\alpha$, and game strategy $s$ for different values of the temptation to defect $T$. All screenshots refer to the prisoner's dilemma game with probabilistic abstention (PDPA) at the $10^5$ Monte Carlo step for a regular square lattice grid populated with $N = 102 \times 102$ agents, with a fixed reward for mutual cooperation $R = 1$, punishment for mutual defection $P = 0$, sucker's payoff $S = 0$ and loner's payoff $L = 0.4$.

Finally, Fig. 4 shows the average fraction of $\varepsilon$ and $\alpha$ on the plane $T - L$ (i.e., temptation to defect vs loner's payoff) for both PDPA and OPD games, with synchronous and asynchronous updating rules. It is possible to observe that the PDPA acts like an enhanced version of the OPD game. In addition, its performance with the asynchronous updating rules is remarkable; we see that when the concept of optionality is given in levels, i.e., the introduction of the probability of abstaining $\alpha$, the population succeeds in controlling the dominance of abstention behaviour, which maintains the diversity of strategies and also helps to promote cooperation.

Furthermore, as discussed previously, despite being more effective in promoting cooperation than the classic games, we observed that cooperation is the dominant strategy only if $T$ is relatively small in a synchronous updating fashion. In summary, for both updating rules, the possibility of not interacting with all neighbours ($\alpha > 0$) helps cooperators to decrease the risk of being exposed to defectors in the initial steps (when most of them could not yet cluster), which consequently allows them to survive even when $T$ is very high. However, this possibility also hampers them from dominating the environment afterwards, which results in the promotion of a robust state of coexistence of both strategies.

## Discussion

We have studied a novel evolutionary game called the prisoner's dilemma with probabilistic abstention (PDPA), which is essentially the merger of two well-known games: the prisoner's dilemma (PD) game and the optional prisoner's dilemma (OPD) game. A number of Monte Carlo simulations with both synchronous and asynchronous updating rules were carried out, where it was shown that the PDPA game is much more beneficial for promoting cooperation than the classic PD and OPD games.

It was discussed that in most evolutionary scenarios (i.e., $T < 1.9$ and $L < 0.8$), the agents who interact at least once ($\alpha < 1$) usually have a better performance. This indicates that intermediate values of $\alpha$ are a better option for promoting both cooperative behaviour and diversity of strategies (cyclic dominance) in the population. Moreover, results suggest that the higher heterogeneity of states in the initial steps play a key role in slowing down the evolution of defection, which increases the chance of the formation of cooperative clusters. It is noteworthy that the precise role of heterogeneity in the PDPA game needs to be further explored. To conclude, it was observed that PDPA is, in fact, an enhanced version of the OPD game, which provides a more realistic representation of the concept of voluntary/optional participation.

## Methods

This work considers the prisoner's dilemma game with probabilistic abstention (PDPA), which is an evolutionary theoretical-game with two pure competing strategies: cooperate (C) and defect (D). In this game, each agent is characterized by two different attributes: game strategy $s$ and the probability of abstaining $\alpha$, which determines how likely it is that an agent will interact in each pairwise play. When an agent abstains from a game interaction, both agents acquire the same loner's payoff $L$. In this way, $\alpha$ is a number from zero to one where $\alpha = 0$ denotes an agent who never abstains (always plays the game), and $\alpha = 1$ denotes an agent who always abstains (never plays the game). When both agents play the game, their payoffs follow the same structure of the classic prisoner's dilemma game, i.e., the reward for mutual cooperation $R = 1$, punishment for mutual defection $P = 0$, $T$ for the temptation to defect and the sucker's payoff $S = 0$. To ensure the proper nature of the dilemma, $1 < T < 2$ and $0 < L < 1$[38].

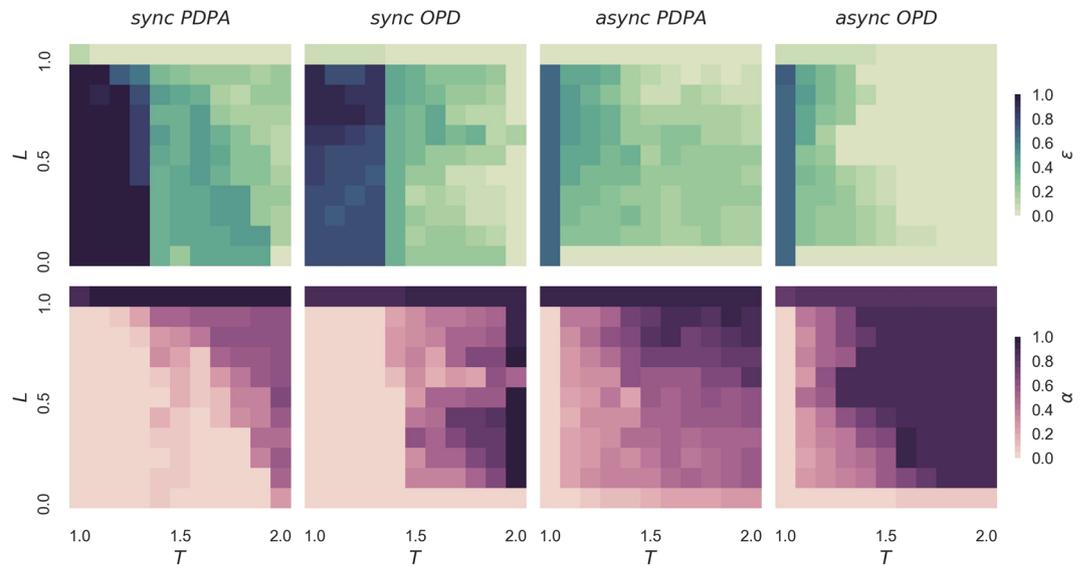

**Figure 4.** Heat map of the average effective cooperation $\varepsilon$ and probability of abstaining $\alpha$ in the $T-L$ plane (i.e., temptation to defect vs loner's payoff) for 100 independent runs at the stationary state (i.e., after $10^5$ Monte Carlo steps). Both synchronous and asynchronous fashions of the optional prisoner's dilemma game (OPD) and the prisoner's dilemma game with probabilistic abstention (PDPA) are explored. A regular square lattice grid is adopted, populated with $N = 102 \times 102$ agents, for a fixed reward for mutual cooperation $R = 1$, punishment for mutual defection $P = 0$, sucker's payoff $S = 0$ and loner's payoff $L = 0.4$.

Without loss of generality, we discretize the values of $\alpha$ to $|\alpha| = 2\kappa$ in equal intervals, where $\kappa$ is the agent's degree. We adopt a regular square lattice grid with periodic boundary conditions (i.e., a toroid) fully populated with $N = 102 \times 102$ agents playing the PDPA game. Each agent interacts with its four immediate neighbours (von Neumann neighborhood) and is initially assigned a strategy $s = \{C, D\}$ and a probability of abstaining $\alpha = \{0, 0.125, 0.250\cdots 0.750, 0.875, 1.0\}$ with equal probability. The evolution process is performed through a number of Monte Carlo (MC) simulations[64] in both synchronous and asynchronous fashion as follows[60]:

- **Synchronous updating:** at each time step, all agents $x$ in the population play the game once with each of their four neighbours $y$ acquiring the payoff $p_{xy}$ for each interaction. After that, for the current time step, each agent copies the strategy and the value of $\alpha$ of the best performing agent in the neighbourhood. In case of ties, or if $x$ is the best in the neighbourhood, its strategy and $\alpha$ remains the same.
- **Asynchronous updating:** at each time step, each agent is selected once on average to play the game and update its strategy and $\alpha$ immediately. That is, in one time step, $N$ agents are randomly selected to perform the following elementary procedures: the agent $x$ plays the game with all neighbours $y$, acquiring the payoffs $p_{xy}$ for each play (i.e., obtaining the utility of $u_x = \sum p_{xy}$); one randomly chosen neighbour of $x$ ($y$) also acquires its payoffs $p_{yz}$ by playing with all its neighbours $z$ (i.e., obtaining the utility of $u_y = \sum p_{yz}$); finally, if $u_y > u_x$, agent $x$ copies the strategy and the value of $\alpha$ from its neighbour $y$ with a probability:

$$W = (1 + exp[(u_x - u_y)/(\kappa K)])^{-1}, \qquad (1)$$

where $K = 0.1$ denotes the amplitude of noise[21].

In our experiments, all Monte Carlo simulations are run for $10^5$ steps, which is a sufficiently long thermalization time to determine the stationary states. Furthermore, to ensure proper accuracy and alleviate the effect of randomness in the approach, the final results are obtained by averaging 100 independent runs.

It is noteworthy that the PDPA game allows us to perform both classic games (PD and OPD). That is, by setting all agents to have $\alpha = 0$, we ensure that they will always play the game, which is essentially the same as considering the classic PD game. Similarly, by setting agents to have $\alpha = \{0, 1\}$ we ensure that some agents will purely abstain, while others will play the game, which is the same as considering the OPD game.

### Acknowledgements

This work was supported by the National Council for Scientific and Technological Development (CNPq-Brazil). Grant number 234913/2014-2.

### Author Contributions

M.C. conceived and conducted the experiments. All authors analysed the results and reviewed the manuscript.

### Additional Information

**Competing Interests:** The authors declare no competing interests.

**Publisher's note:** Springer Nature remains neutral with regard to jurisdictional claims in published maps and institutional affiliations.